\documentclass[twocolumn,prl,showpacs,preprintnumbers]{revtex4}
\usepackage{amssymb}
\usepackage{amsmath}
\usepackage{graphicx}
\usepackage{dcolumn}
\usepackage{bm}

\begin{document}

\title{Entanglement Monogamy of Tripartite Quantum States}
\author{Chang-shui Yu}
\author{He-shan Song}
\email{hssong@dlut.edu.cn}
\affiliation{School of Physics and Optoelectronic Technology, Dalian University of
Technology, Dalian 116024, P. R. China}
\date{\today }

\begin{abstract}
An interesting monogamy equation with the form of Pythagorean
theorem is found for $2\otimes 2\otimes n$-dimensional pure states,
which reveals the relation among bipartite concurrence, concurrence
of assistance, and genuine tripartite entanglement. At the same
time, a genuine tripartite entanglement monotone as a generalization
of 3-tangle is naturally obtained for $(2\otimes 2\otimes n)$-
dimensional pure states in terms of a distinct idea. For mixed
states, the monogamy equation is reduced to a monogamy inequality.
Both results for tripartite quantum states can be employed to
multipartite quantum states.
\end{abstract}

\maketitle
\section{I. Introduction}
Entanglement is an essential feature of quantum mechanics, which
distinguishes quantum from classical world. A key property of
entanglement as well as one of the fundamental differences between
quantum entanglement and classical correlations is the degree of
sharing among many parties -----Unlike classical correlations,
quantum entanglement is monogamous [1-3], i.e., the degree to which
either of two parties can be entangled with anything else seems to
be constrained by the entanglement that may exist between the two
quantum parties. For the systems of three qubits, a kind of monogamy
of bipartite quantum entanglement measured by concurrence [4] was
described by Coffman-Kundu-Wootters (CKW) inequality [1]. The
generalization to the case of multiple qubits was conjectured by CKW
and has been proven recently by Osborne et al [5]. The monogamy
inequality dual to CKW inequality based on concurrence of assistance
(CoA) [6] was presented for tripartite systems of qubits by Gour et
al [7] and the generalized one for multiple qubits was proven in
Ref. [8]. In this paper, we find a new and
very interesting monogamy equation for $\left( 2\otimes 2\otimes n\right) $%
-dimensional (or multiple qubits) quantum pure states which relates the
bipartite concurrence, CoA and genuine tripartite entanglement.

In fact, CKW inequality and the dual one correspond to a residual
quantity, respectively. It is only for tripartite pure states of
qubits that so far the two residual quantities have been shown to be
the same and have clear physical meanings. From Ref. [1] and [6],
one can learn that it just corresponds to 3-tangle [1]. One of the
distinguished advantages of our monogamy equation will be found that
the residual quantity has clear
physical meanings not only for $\left( 2\otimes 2\otimes n\right) $%
-dimensional quantum pure state but also for a general multipartite
pure state including a pair of qubits.

Recently it has been realized that entanglement is a useful physical
resource for various kinds of quantum information processing [9-12].
Based on the different physics of implementation, there are usually
three alternative ways [7] to producing entanglement. The specially
important way for quantum communication is the reduction of a
multipartite entangled state to an entangled state with fewer
parties, which is called "assisted entanglement" quantified by
entanglement of assistance (EoA) [13]. An important application of
EoA is for tripartite quantum entangled state to maximize the
entanglement of two parties (qubits) denoted by Alice and Bob with
the assistance of the third party (qudit) named Charlie who is only
allowed to do local operations. However, because EoA is not an
entanglement monotone [14], one would prefer to the remarkable
entanglement monotone------concurrence of assistance (CoA) where
concurrence is employed to quantify the entanglement between Alice
and Bob. In this process of entanglement preparation, Charlie only
makes local operations and classical communications in order to
increase the entanglement shared by Alice and Bob, therefore it is
impossible to produce new entanglement. There must exist some
trade-off between the increment of entanglement shared by Alice and
Bob induced by Charlie and quantum correlations with other forms.
Then what are those?

The question is answered in this paper by our interesting monogamy
equation. From the equation, one can find that the increment of
entanglement shared by Alice and Bob just corresponds to the degree
of genuine tripartite entanglement (3-way entanglement) of $\left(
2\otimes 2\otimes n\right) $- dimensional quantum pure state and is
analytically calculable. Hence, the increment naturally
characterizes the genuine tripartite entanglement, which is shown to
be an entanglement monotone and can be considered as an interesting
generalization of 3-tangle in terms of a new idea. In addition, the
monogamy equation is reduced to a monogamy inequality for mixed
states. The results are also suitable for multipartite quantum
states. This paper is organized as follows. We first introduce our
interesting monogamy equation for pure states; Then for mixed
states, we reduce this monogamy equation for pure state to a
monogamy inequality; Next we point out these results are suitable
for multipartite quantum states; The conclusion is drawn finally.

\section{II. Monogamy equation for pure states}
Given a tripartite
$\left( 2\otimes 2\otimes n\right) $- dimensional quantum pure state
$\left\vert \Psi \right\rangle _{ABC}$ shared by three parties
Alice, Bob and Charlie, where Charlie's aim is to maximize the
entanglement shared by Alice and Bob by local measurements on
Charlie's particle C, the reduced density matrix by tracing over
party C can be given by $\rho _{AB}=Tr_{C}$ $\left( \left\vert
\Psi \right\rangle _{ABC}\left\langle \Psi \right\vert \right) $. Let $%
\mathcal{E}=\{p_{i},\left\vert \varphi _{i}^{AB}\right\rangle \}$ is any a
decomposition of $\rho _{AB}$ such that
\begin{equation}
\rho _{AB}=\sum\limits_{i}p_{i}\left\vert \varphi _{i}^{AB}\right\rangle
\left\langle \varphi _{i}^{AB}\right\vert ,\sum\limits_{i}p_{i}=1,
\end{equation}%
then CoA is defined [5,6] by
\begin{eqnarray}
C_{a}\left( \left\vert \Psi \right\rangle _{ABC}\right) &=&\max_{\mathcal{E}%
}\sum\limits_{i}p_{i}C\left( \left\vert \varphi _{i}^{AB}\right\rangle
\right) \\
=C_{a}\left( \rho _{AB}\right) &=&tr\sqrt{\sqrt{\rho _{AB}}\tilde{\rho}_{AB}%
\sqrt{\rho _{AB}}} \\
&=&\sum\limits_{i=1}^{4}\lambda _{i},
\end{eqnarray}%
where $\tilde{\rho}_{AB}=\left( \sigma _{y}\otimes \sigma
_{y}\right) \rho _{AB}^{\ast }\left( \sigma _{y}\otimes \sigma
_{y}\right) $, $\sigma_y$ is Pauli matrix and
\begin{equation}
C\left( \rho _{AB}\right) =\max \{0,\lambda _{1}-\sum\limits_{i>1}\lambda
_{i}\}
\end{equation}
is the concurrence of the reduced density matrix $\rho _{AB}$ with $\lambda
_{i}$ being the square roots of the eigenvalues of $\rho _{AB}\tilde{\rho}%
_{AB}$ in decreasing order. With the definitions of CoA and
concurrence, we can obtain the following theorem.

\textbf{Theorem 1: } \emph{For a }$\left( 2\otimes 2\otimes n\right) $\emph{%
- dimensional quantum pure state }$\left\vert \Psi \right\rangle _{ABC}$%
\emph{, }%
\begin{equation}
C_{a}^{2}\left( \rho _{AB}\right) =C^{2}\left( \rho _{AB}\right) +\tau
^{2}\left( \rho _{AB}\right) ,
\end{equation}%
\emph{\newline
where }$\tau \left( \rho _{AB}\right) =\tau \left( \left\vert \Psi
\right\rangle _{ABC}\right) $\emph{\ is the genuine tripartite entanglement
measure for }$\left\vert \Psi \right\rangle _{ABC}$\emph{.}

It is very interesting that eq. (6) has an elegant form that is
analogous to Pythagorean theorem if one considers CoA as the length
of the hypotenuse of a right-angled triangle and considers bipartite
concurrence and genuine tripartite entanglement as the lengths of
the other two sides of the triangle. Note that the lengths of all
the sides are allowed to be zero. The illustration of the relation
is shown in Fig. 1 (See the left triangle).

\textbf{Proof.} According to the definition of CoA and concurrence, it is
obvious that
\begin{equation}
C_{a}^{2}\left( \rho _{AB}\right) -C^{2}\left( \rho _{AB}\right) \geq 0.
\end{equation}%
Then the remaining is to prove that $\tau \left( \rho _{AB}\right) $
is an entanglement monotone and characterizes the genuine tripartite
entanglement of $\left\vert \Psi \right\rangle _{ABC}$. Next, we
first prove that $\tau \left( \rho _{AB}\right) $ does not increase
under a general tripartite
local operation and classical communication (LOCC) denoted by $\mathcal{M}%
_{k}$ where subscript $k$ labels different outcomes. We first assume
that Alice and Bob perform quantum operations $M_{Akj}$, and
$M_{Bkj}$ on their qubits respectively, where $\sum\limits_{k,j}$
$M_{Akj}^{\dagger }M_{Akj}\leq I_{A}$ and $\sum\limits_{k,j}$
$M_{Bkj}^{\dagger }M_{Bkj}\leq I_{B}$ are the most general local
operations given in terms of the Kraus operator [15] with $I_{A}$
and $I_{B}$ being the identity operators in Alice's and Bob's
systems. After local operations, the average CoA can be written as
\begin{eqnarray}
&&\sum\limits_{kk^{\prime }}P_{kk^{\prime }}\tau \left( \mathcal{M}%
_{kk^{\prime }}\left( \rho _{AB}\right) \right)  \notag \\
&=&\sum\limits_{kk^{\prime}}P_{kk^{\prime }}\sqrt{C_{a}^{2}\left( \mathcal{M}%
_{kk^{\prime }}\left( \rho _{AB}\right) \right) -C^{2}\left( \mathcal{M}%
_{kk^{\prime }}\left( \rho _{AB}\right) \right) }  \notag \\
&\leq &\left\{ \left[ \sum\limits_{kk^{\prime}}P_{kk^{\prime}}C_{a}\left( \mathcal{M}%
_{kk^{\prime }}\left( \rho _{AB}\right) \right) \right] ^{2}\right.
\notag
\\
&&\left. -\left[ \sum\limits_{kk^\prime}P_{kk^{\prime}}C\left(
\mathcal{M}_{kk^{\prime
}}\left( \rho _{AB}\right) \right) \right] ^{2}\right\} ^{1/2}  \notag \\
&=& \sum\limits_{kk^{\prime }jj^{\prime }}\left\vert \det \left(
M_{Akj}\right) \det \left( M_{Bk^{\prime }j^{\prime }}\right) \right\vert
\sqrt{C_{a}^{2}\left( \rho _{AB}\right) -C^{2}\left( \rho _{AB}\right) }
\notag \\
&\leq &\sqrt{C_{a}^{2}\left( \rho _{AB}\right) -C^{2}\left( \rho
_{AB}\right) }=\tau \left( \rho _{AB}\right) ,
\end{eqnarray}%
where
\begin{eqnarray}
&&\mathcal{M}_{kk^{\prime }}\left( \rho _{AB}\right)
=\sum\limits_{jj^{\prime }}\left( M_{Akj}\otimes M_{Bk^{\prime
}j^{\prime
}}\otimes I_{C}\right)  \notag \\
&\times& \left\vert \Psi \right\rangle _{ABC}\left\langle \Psi
\right\vert \left( M_{Akj}^{\dagger }\otimes M_{Bk^{\prime
}j^{\prime }}^{\dagger }\otimes I_{C}\right) /P_{kk^\prime},
\end{eqnarray}%
and $P_{kk^{\prime }}=tr\mathcal{M}_{kk^{\prime }}\left( \left\vert
\Psi \right\rangle _{ABC}\left\langle\Psi\right\vert\right) $. Here
the first inequality follows from Cauchy-Schwarz inequality:
\begin{equation}
\sum\limits_{i}x_{i}y_{i}\leq \left( \sum\limits_{i}x_{i}^{2}\right)
^{1/2}\left( \sum\limits_{j}y_{j}^{2}\right) ^{1/2},
\end{equation}%
the second inequality follows from the geometric-arithmetic inequality
\begin{equation}
\sum\limits_{kj}\left\vert \det \left( M_{xkj}\right) \right\vert \leq \frac{%
1}{2}\sum\limits_{kj}trM_{xkj}^{\dagger }M_{xkj}\leq 1,x=A,B,
\end{equation}%
and the second equation is derived from the fact [4,16] that
\begin{equation}
C_{a}\left( M_{Akj}\rho _{AB}M_{Akj}^{\dagger }\right) =\left\vert \det
\left( M_{Akj}\right) \right\vert C_{a}\left( \rho _{AB}\right) ,
\end{equation}%
\begin{equation}
C\left( M_{Akj}\rho _{AB}M_{Akj}^{\dagger }\right) =\left\vert \det \left(
M_{Akj}\right) \right\vert C\left( \rho _{AB}\right)
\end{equation}%
and the analogous relations for $M_{Bk^{\prime }j^{\prime }}$. Eq. (8) shows
that $\tau \left( \rho _{AB}\right) $ does not increase under Alice's and
Bob's local operations.
\begin{figure}[tbp]
\centering
\includegraphics[width=8.5cm]{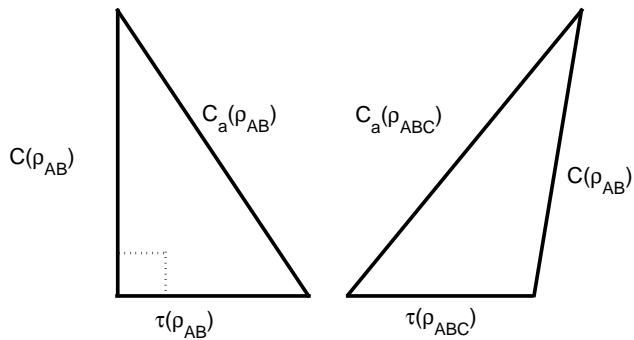}
\caption{The illustration of the relation among CoA, bipartite
concurrence and genuine tripartite entanglement. The left
right-angled triangle corresponds to Theorem 1 (for pure states) and
the right obtuse-angled triangle corresponds to Theorem 2 (for mixed
states). All the quantities given in the figures are defined the
same as the corresponding theorems.}
\end{figure}

Next we prove that $\tau \left( \rho _{AB}\right) $ does not
increase under Charlie's local operations either. Suppose $\rho
_{AB}=\lambda \rho
_{1}^{AB}+(1-\lambda )\rho _{2}^{AB}$, $\lambda \in \lbrack 0,1]$, then%
\begin{eqnarray}
&&\lambda \tau \left( \rho _{1}^{AB}\right) +(1-\lambda )\tau \left( \rho
_{2}^{AB}\right)   \notag \\
&=&\lambda \sqrt{C_{a}^{2}\left( \rho _{1}^{AB}\right) -C^{2}(\rho _{1}^{AB})%
}  \notag \\
&&+(1-\lambda )\sqrt{C_{a}^{2}\left( \rho _{2}^{AB}\right) -C^{2}(\rho
_{2}^{AB})}  \notag \\
&\leq &\left\{ \left[ \lambda C_{a}\left( \rho _{1}^{AB}\right) +(1-\lambda
)C_{a}\left( \rho _{1}^{AB}\right) \right] ^{2}\right.   \notag \\
&&\left. -\left[ \lambda C\left( \rho _{2}^{AB}\right) +(1-\lambda )C\left(
\rho _{2}^{AB}\right) \right] ^{2}\right\} ^{1/2}  \notag \\
&\leq &\sqrt{C_{a}^{2}\left( \rho _{AB}\right) -C^{2}\left( \rho
_{AB}\right) }=\tau \left( \rho _{AB}\right) ,
\end{eqnarray}%
where the first inequality follows from the Cauchy-Schwarz
inequality (10) and the second inequality follows from the
definitions of $C_{a}\left( \rho _{AB}\right) $ and $C\left( \rho
_{AB}\right) $. Eq. (14) shows that $\tau \left( \rho _{AB}\right) $
is a concave function of $\rho _{AB}$, which proves that $\tau
\left( \rho _{AB}\right) $ does not increase under
Charlie's local operations following the same procedure (or \textbf{Theorem 3%
}) in Ref. [17]. All above show that $\tau \left( \rho _{AB}\right)
$ is an entanglement monotone.

Now we prove that $\tau\left(\rho_{AB}\right)$ characterizes genuine
tripartite entanglement. Based on eq. (4) and eq. (5), it is obvious
that
\begin{equation}
\tau \left( \rho _{AB}\right) =\left\{
\begin{array}{cc}
\sum\limits_{i=1}^{4}\lambda _{i}, & \lambda _{1}\leq
\sum\limits_{i=2}^{4}\lambda _{i}, \\
2\sqrt{\lambda _{1}\sum\limits_{i=2}^{4}\lambda _{i}}, & \lambda
_{1}>\sum\limits_{i=2}^{4}\lambda _{i},%
\end{array}%
\right.
\end{equation}%
which is an explicit formulation. Ref. [18] has given a special
quantity named "entanglement semi-monotone" that characterizes the
genuine tripartite entanglement. One can find that it requires the
same conditions as the quantity introduced in Ref. [18] for $\tau
\left( \rho _{AB}\right) $ to reach \emph{zero}, which shows that
$\tau \left( \rho _{AB}\right) $ characterizes the genuine
tripartite entanglement. The proof is completed.$\hfill \Box $

In general, multipartite entanglement is quantified in terms of
different classifications [19-21]. However, $\tau \left( \rho
_{AB}\right)$ quantifies genuine tripartite entanglement in a new
way, i.e., we consider the entanglement of GHZ-state class as the
minimal unit [22] in terms of tensor treatment [23] and summarize
all the genuine tripartite inseparability without further
classifications. It is an interesting generalization of 3-tangle.
\textbf{Theorem 1} shows a very clear physical meaning, i.e. the
increment of entanglement between Alice and Bob induced by Charlie
is just the genuine tripartite entanglement among them. The meaning
can especially easily be understood for tripartite quantum state of
qubits. In this case, $\tau \left( \rho _{AB}\right)
=2\sqrt{\lambda_1\lambda_2}$. Two most obvious examples are GHZ
state and W state. The entanglement of reduced density matrix of GHZ
state is zero, hence \textbf{Theorem 1} shows that the CoA of GHZ
state all comes from the three-way entanglement and equals to 1 (the
value of 3-tangle). On the contrary, the W state has no three-way
entanglement (only two-way entanglement) [24], hence its CoA is only
equal to the concurrence ($\frac{2}{3}$) of two parties. That is to
say, for W state, Charlie can not provide any help to increase the
entanglement between Alice and Bob.

\section{III. Monogamy inequality for mixed states}
For a given mixed state $\rho _{ABC}$, CoA can be extended to mixed
states in terms of convex roof construction [15], i.e.,
\begin{equation}
C_{a}(\rho _{ABC})=\min \sum\limits_{i}p_{i}C_{a}(\left\vert \psi
^{i}\right\rangle _{ABC}),
\end{equation}%
where the minimum is taken over all decompositions $\{p_{i},\left\vert \psi
\right\rangle _{ABC}\}$ of $\rho _{ABC}$. Thus we have the following theorem.

\textbf{Theorem 2.}-\emph{For a }$\left( 2\otimes 2\otimes n\right)
$\emph{-
dimensional mixed state }$\rho _{ABC}$\emph{,}%
\begin{equation}
C_{a}^{2}\left( \rho _{ABC}\right) \geqslant C^{2}\left( \rho _{AB}\right)
+\tau ^{2}\left( \rho _{ABC}\right) ,
\end{equation}%
\emph{where }$\tau \left( \rho _{ABC}\right) $\emph{\ is the genuine
tripartite entanglement measure for mixed states by extending }$\tau
\left( \cdot\right) $\emph{\ of pure states in terms of convex roof
construction and }$\rho _{AB}=tr_{C}\rho _{ABC}$\emph{.}

Analogous to \textbf{Theorem 1}, one can easily find that the
relation of \textbf{Theorem 2} corresponds to an obtuse-angled
triangle after a simple algebra, where CoA corresponds to the length
of the side opposite to the obtuse angle. See the right triangle in
Fig.1 for the illustration.

\textbf{\ Proof}. Suppose $\{p_{k},\left\vert \psi ^{k}\right\rangle
_{ABC}\} $ is the optimal decomposition in the sense of

\begin{equation}
\tau \left( \rho _{ABC}\right) =\sum\limits_{k}p_{k}\tau \left( \left\vert
\psi ^{k}\right\rangle _{ABC}\right) =\sum\limits_{k}p_{k}\tau \left( \sigma
_{AB}^{k}\right) ,
\end{equation}%
where $\sigma _{AB}^{k}=tr_{C}\left[ \left\vert \psi ^{k}\right\rangle
_{ABC}\left\langle \psi ^{k}\right\vert \right] $. According to \textbf{%
Theorem 1}, we have
\begin{eqnarray}
\tau \left( \rho _{ABC}\right) &=&\sum\limits_{k}p_{k}\sqrt{C_{a}^{2}\left(
\sigma _{AB}^{k}\right) -C^{2}\left( \sigma _{AB}^{k}\right) } \notag\\
&\leq &\sqrt{\left[ \sum\limits_{k}p_{k}C_{a}\left( \sigma _{AB}^{k}\right) %
\right] ^{2}-\left[ \sum\limits_{k}p_{k}C\left( \sigma _{AB}^{k}\right) %
\right] ^{2}}  \notag \\
&\leq &\sqrt{C_{a}^{2}\left( \rho _{ABC}\right) -C^{2}\left( \rho
_{AB}\right) },
\end{eqnarray}%
where the first inequality follows from Cauchy-Schwarz inequality
(10) and the second inequality holds based on the definitions of
$C_{a}\left( \rho _{AB}\right) $ and $C\left( \rho _{AB}\right) $.
Eq. (19) finishes the proof. $\hfill \Box $

\section{IV. Monogamy for multipartite quantum states}
 Any a given
$N$-partite quantum state can always be considered as a $\left(
2\otimes 2\otimes X \right)$- dimensional tripartite quantum states
with $X$ denoting the total dimension of $N-2$ subsystems so long as
the state includes at least two qubits, hence both the two theorems
hold in these cases. However, it is especially worthy of being noted
that the two qubits must be owned by Alice and Bob respectively and
the other $N-2$ subsystems should be at Charlie's hand and be
considered as a whole. Charlie is allowed to perform any nonlocal
operation on the $N-2$ subsystems. In addition, there may be
different groupings [25] of a multipartite quantum state especially
for multipartite quantum states of qubits, hence there exist many
analogous monogamy equations (for pure states) or monogamy
inequalities (for mixed states) for the same quantum state. For pure
states, every monogamy equation will lead to a genuine $\left(
2\otimes 2\otimes X\right) $- dimensional tripartite entanglement
monotone that quantifies the genuine tripartite entanglement of the
tripartite state generated by the corresponding grouping.

\section{V. Conclusion and discussion}We have presented an
interesting monogamy equation with elegant form for $(2\otimes
2\otimes n)$- dimensional quantum pure states, which, for the first
time, reveals the relation among bipartite concurrence, CoA and
genuine tripartite entanglement. The
equation naturally leads to a genuine tripartite entanglement measure for $%
(2\otimes 2\otimes n)$-dimensional tripartite quantum pure states,
which quantifies tripartite entanglement in terms of a new idea. The
monogamy equation can be reduced to a monogamy inequality for mixed
states. Both the results for tripartite quantum states are also
suitable for multipartite quantum states. We hope that the current
results can shed new light on not only the monogamy of entanglement
but also the quantification of multipartite entanglement.

\section{ Acknowledgement} This work was supported by the National
Natural Science Foundation of China, under Grant No. 10747112 and
No. 10575017.


\begin{thebibliography}{99}
\bibitem{[22]} Ref. [19,20] have introduced onionlike classification of
tripartite entanglement for $2\otimes 2\otimes n$- dimensional
quantum pure states. It has been shown that the quantum states in
the outer class can always irreversely converted into the states in
the inner class. The GHZ-state class belongs to the most inner class
that characterizes genuine tripartite entanglement (three-way
entanglement). Thus entanglement of GHZ-state class can be
understood as the minimal unit that quantifies genuine tripartite
inseparability.
\end{thebibliography}
\end{document}